\documentclass[journal]{IEEEtran}
\usepackage{graphicx,times,amsmath,booktabs,algorithm,algorithmic,amsmath}
\usepackage{multicol}
\usepackage{amsmath,amssymb}
\usepackage{subfigure}
\usepackage{stfloats}
\usepackage{setspace} 
\usepackage{color}

\begin{document}
\title{Air-Ground Integrated Mobile Edge Networks: \\Architecture, Challenges and Opportunities}

\author{Nan~Cheng,
        Wenchao~Xu,
        Weisen~Shi,
        Yi~Zhou,
        Ning~Lu,
        Haibo~Zhou,
        Xuemin (Sherman) Shen\\

\thanks{N. Cheng, W. Xu, W. Shi, and X. Shen are with the Electrical and Computer Engineering Department, University of Waterloo, Waterloo, ON, N2L3G1, Canada (emails: \{n5cheng,w74xu,w46shi,sshen\}@uwaterloo.ca).}
\thanks{Y. Zhou is with the School of Computer and Information Engineering, Henan University, Kaifeng, Henan, China  (email: zhouyi@henu.edu.cn). Y. Zhou is the corresponding author of the article.}
\thanks{N. Lu is with the Department of Computing Science, Thompson Rivers University, Kamloops, BC V2C 0C8, Canada  (email: nlu@tru.ca).}
\thanks{H. Zhou is with the School of Electronic Science and Engineering, Nanjing University, Nanjing, Jiangsu, China  (email: haibozhouuw@gmail.com).}
 }

\maketitle
\begin{abstract}
The ever-increasing mobile data demands have posed significant challenges in the current radio access networks, while the emerging computation-heavy Internet of things (IoT) applications with varied requirements demand more flexibility and resilience from the cloud/edge computing architecture. In this article, to address the issues, we propose a novel air-ground integrated mobile edge network (AGMEN), where UAVs are flexibly deployed and scheduled, and assist the communication, caching, and computing of the edge network. In specific, we present the detailed architecture of AGMEN, and investigate the benefits and application scenarios of drone-cells, and UAV-assisted edge caching and computing. Furthermore, the challenging issues in AGMEN are discussed, and potential research directions are highlighted.
\end{abstract}
\IEEEpeerreviewmaketitle


\section{Introduction}
With the fast development of the mobile Internet, the data traffic has witnessed an exponential growth over the last few years. 
In particular, the emerging data-craving applications, such as high-quality video streaming, virtual reality (VR), augmented reality (AR), and remote-operated driving, will pose strict requirements on the communication and computation capabilities of the network. Meanwhile, the dramatically increasing mobile devices and massive connections of machine-type communications (MTC) in Internet of Things (IoT) demand a full support of massive connections with low latency, high reliability, and low power consumption. Many efforts have been done in advanced cellular networks and cloud computing to satisfy such requirements. In recent standardization of 5G cellular networks, to achieve extremely high data rates and massive MTC, advanced technologies have been suggested, such as Polar codes, massive MIMO, full-duplex radio, millimeter-wave (mm-Wave) communication, beamforming, and so forth. With mobile cloud computing (MCC), the wireless devices can offload the computation-intensive tasks to the cloud servers with abundant computational resources to improve the task processing efficiency and reduce the user hardware costs, with the consideration of energy consumption, user mobility and context information, and communication delay and cost. However, the base station centric cellular networks may still suffer from the traffic overloading, and can hardly satisfy the exponential growing mobile traffic and the various new requirements of the emerging applications. In addition, the traditional cloud computing faces the problems of long response delay and limited backhaul bandwidth. For example, in image processing applications, a large batch of high quality images or pictures should be uploaded from the user devices to the distant cloud servers. This could pose severe burden on the backhaul network and bring excessive delay due to the long communication distance.

\begin{figure*}
\centering
\includegraphics[height=7cm]{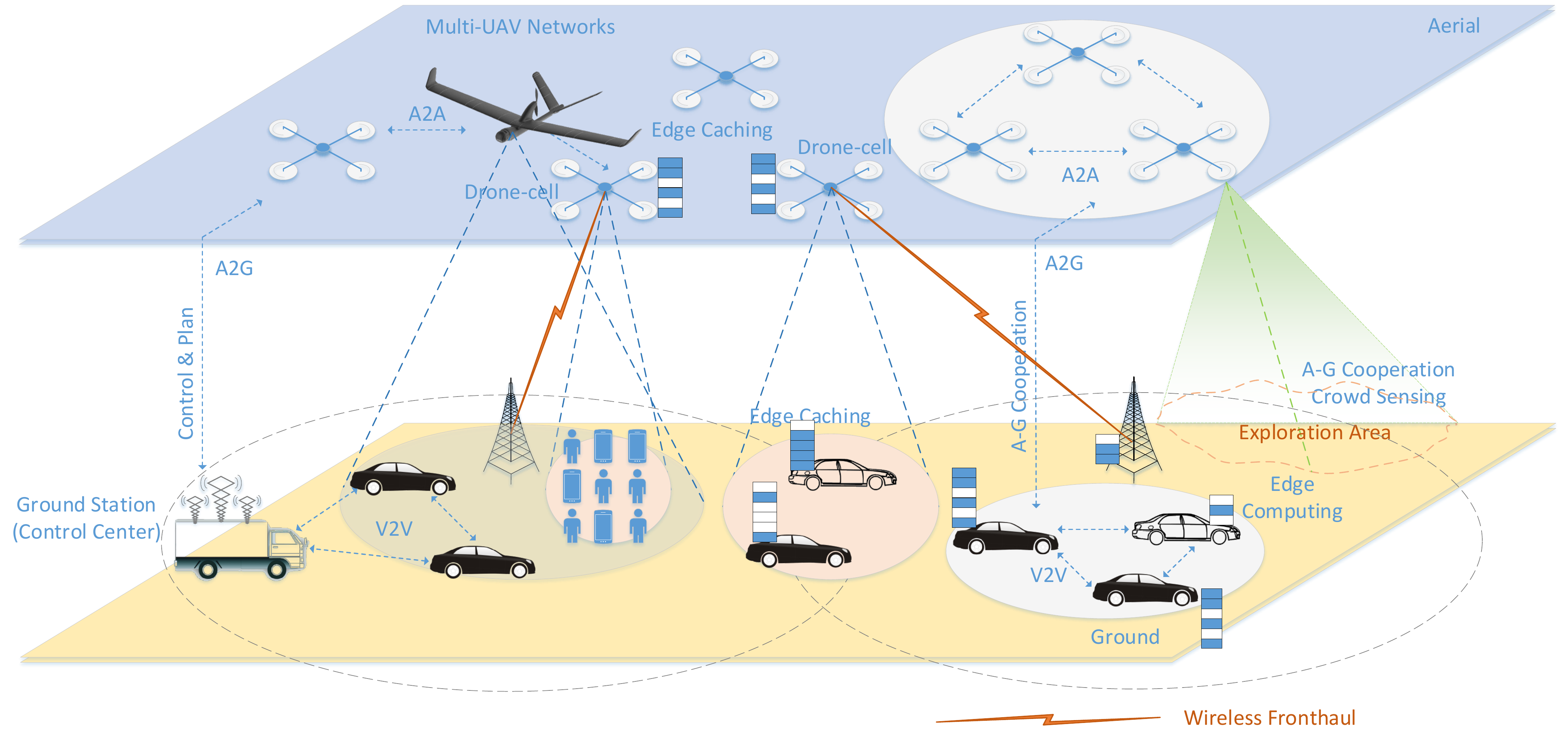}
\caption{An overview of AGMEN architecture.}
\label{fig:archi}
\end{figure*}

Mobile edge network (MEN) is a promising solution to address the above issues. By moving the network functions and resources closer to the end users, i.e., the network edge, many benefits can be obtained, such as high data rates, low delay, improved energy efficiency, and flexible network deployment and management \cite{wang2017survey}. In terms of the functionalities of the MEN, i.e., communication, caching, and computing, the main MEN technologies include network densification, mobile edge caching, and mobile edge computing.
\begin{enumerate}
\item[$\bullet$] \emph{Network densification}. Ultra-dense network is a key component of the 5G network. The communication resources are moving to small-size cells which are densely deployed and closer to the network edge, such as pico cells, femto cells, and high-efficiency WiFi networks, forming the heterogeneous networks (HetNets). The Ultra-dense network can not only reap the benefits of high quality links such that high-frequency spectrum bands, such as mm-Wave communication, can be employed, but also harness more spatial reuse gain to boost the network capacity \cite{qian2017dynamic}. In spite of the benefits such as high data rate and network capacity, and low energy consumption, network densification also introduces some challenging issues, e.g., interference and handover management. In addition, the backhaul bandwidth constrains can limit the performance of HetNets.
\item[$\bullet$] \emph{Mobile edge caching}. Due to the increased capacity and reduced cost of the storage devices, network caching schemes are widely employed to offload the burden of the backhaul network. Since certain popular contents (e.g., on-the-air TV series and popular music) are frequently requested, such contents can be cached during off-peak times in the network edge, such as base stations and even user devices \cite{su2017edge}. Then, the contents are distributed to the requesters through the high-rate and low-cost mobile edge networks, rather than transmitted through the backhaul network repeatedly. In addition, proximate communication, such as device-to-device (D2D) communication can be used to share the user cached contents among nearby user devices efficiently.
\item[$\bullet$] \emph{Mobile edge computation}. In mobile edge computing, the computing servers are deployed in edge nodes, such as base stations and user devices with high computing capability \cite{tao2017performance,liu2018multiobjective}. Therefore, the computation-intensive tasks, such as VR and image processing, can be offloaded to mobile edge servers instead of the long-distance cloud servers. Mobile edge computation can offer benefits such as high bandwidth, low latency, low cost, and abundant user and context information. However, the user mobility becomes a challenging issue in the computing task offloading decision making in mobile edge computation.
\end{enumerate}

The VehiculAr NETworks (VANETs) have significantly benefited from the concept of MEN. The densely deployed small cells offload the heavy vehicular mobile traffic, providing a cost-effective mobile Internet access at vehicular speeds \cite{cheng2016opportunistic}. Employing vehicle mobility prediction, contents can be pre-fetched and cached in small cell base stations, WiFi APs, and vehicle on-board storage devices, and distributed using short-range cost-effective communication paradigms, such as D2D communications and delay-tolerant networks \cite{cheng2017performance}. Furthermore, the future car will be outfitted with car AI computers for computing tasks such as self-driving, and thus the computation resources of vehicles can be shared within the MEN to improve the resource utilization. However, the vehicular MEN has certain limitations and problems. The mobile traffic demands of vehicular networks varies greatly temporally and spatially, due to high mobility and changing density of vehicles, as well as the varied QoS requirements of the mobile services. In VANETs, the network connectivity might be degraded in special cases, such as rural roads with sparse vehicles. In addition, it is challenging to schedule computation task offloading among vehicle considering the high mobility. Therefore, a rigidly deployed MEN will have difficulty in dealing with the dynamic communication and computation demands of VANETs.

Recently, the unmanned aerial vehicles (UAVs), especially mini-UAVs, have attracted much attention due to the flexible deployment, agile management, and low cost, and have been widely used in military and civil applications. Recent research has focused on employing UAV communication to assist the ground networks. An extensive survey of UAV communication networks can be found in \cite{gupta2016survey}. In \cite{zhou2015multi}, the aerial-ground cooperative VANETs is studied, where the architecture and key challenges are discussed. In terms of MEN, the UAV can provide radio access to a group of users, which is studied in \cite{bor2016new}, while the content caching in UAVs for edge users is investigated in \cite{chen2017caching}, and UAV-assisted edge computing for IoT services is studied in \cite{jeong2017mobile}. However, there is no comprehensive architecture of an air-ground integrated MEN, where drone-cells, UAV-assisted edge caching and computing are jointly designed and optimized.

To address the issues in both air and ground networks and integrate UAV-assisted network densification, edge caching and computing, we propose a novel architecture of Aerial-Ground Integrated Mobile Edge Network (AGMEN). In the proposed AGMEN, multiple drone-cells are deployed in a flexible manner to provide agile radio access network (RAN) coverage for the temporally and spatially changing users and data traffic. The exponentially growing storing and computing capabilities of vehicles, especially self-driving vehicles, are employed to fulfill the mobile caching and computing tasks, while the UAVs can serve as edge network controller to efficiently allocate computing and storage resources. Equipped with IoT devices such as cameras and sensors, UAVs can conduct specific computing tasks or serve as fog computing platform for various IoT services. Also, UAVs can cache popular contents for reducing the burden of fronthaul and backhaul networks. In this article, we will elaborate the novel AGMEN architecture, present the benefits of UAV-assisted network functions, discuss the challenges and potential solutions, and highlight promising research directions for future study.

\section{Air-Ground Integrated Mobile Edge Networks}
In this section, we propose the novel AGMEN architecture to facilitate the air-ground cooperation. We first describe the AGMEN architecture, where an air-ground two-layer cooperative networking is introduced and explained. Then, we present the crucial components of AGMEN, i.e., multi-access RAN, edge caching, and edge computing, and discuss key functions of each component.

\subsection{AGMEN Architecture}
The overall architecture of AGMEN is shown in Fig. \ref{fig:archi}. The AGMEN has a two-layer networking architecture, where UAVs are deployed to set up a multi-UAV aerial network, and mobile users, vehicles, and RAN infrastructure form the ground network. In the aerial network, UAVs are outfitted with sensors, communication modules, embedded processors, and storage devices, through which the UAVs can behave as a multi-functional network controller. Via aerial-to-aerial (A2A) communications using heterogeneous radio interfaces, such as IEEE 802.15.4 or WiFi, UAVs can transmit information among each other, including sensing data, control and coordination information, form a flying ad hoc network (FANET). Through the information exchange and controlled mobility, the aerial network can conduct specific tasks, such as maintaining the connectivity of a wireless sensor network, date-mule for delay and disruption tolerant networks (DTNs), traffic monitoring and remote sensing, and so forth. In the ground network, the heterogeneous RAN, including macro cells, small cells, and WiFi, serve the mobile users, such as mobile phones, self-driving cars, IoT devices, etc. Cooperation between aerial network and ground network is enabled through aerial-to-ground (A2G) communications. The mobility of UAVs can be controlled by the ground control center, and the data collected by the aerial network is sent to the ground center for further processing and utilization. In addition, since the UAVs have a better sensing scope than ground users due to the high altitude, UAVs can first conduct a large-scale sensing, and guide the ground users for fine-grained sensing or rescue tasks.

Reaping the characteristics of UAVs and the two-layer network architecture, many benefits can be brought to AGMEN. By employing the wireless fronthaul connection, the UAVs can serve as the BSs for the small cells, named drone-cells, as shown in Fig. \ref{fig:archi}, providing flexible Internet access for a group of ground users. Popular contents can be cached in the UAVs or ground vehicles, and transmitted through the AGMEN, e.g., the drone-cells or D2D communication among users. In addition, the exponentially increasing computing capability of vehicles can be employed for edge computing, where UAVs can schedule the computing tasks while on-vehicle computers fulfill the tasks. In the sequel, the AGMEN functionalities will be described from the perspective of UAV drone-cell, UAV-assisted edge caching, and UAV-based edge computing for IoT services, respectively.

\subsection{Multi-Access RAN with Drone-Cells}
In the LTE networks and currently developing 5G networks, network densification is envisioned as a potential solution to the significantly growing mobile data demands. Multi-access HetNets consisting of macro-cells and many kinds of small cells can greatly increase the capacity and improve the energy efficiency of the network. However, such fixed deployed HetNets will be straining to satisfy the dynamic mobile traffic of the future networks, especially when the flourishing IoT devices and services are considered. Due to the temporal and spatial variety of the data traffic, the network may be overloaded handling a burst geographical traffic, while the network resources are wasted when little traffic happens.
UAVs, serving as aerial BSs, can provide drone-cell coverage to IoT devices and mobile users. Multi-access drone-cells are envisioned to provide extra flexibility and manageability to deal with the dynamic traffic demands. The drone-cells can be dispatched to specific areas in designed time periods to offer cost-effective radio access to heavy data traffic. For example, in \cite{yang2017proactive}, a flexible drone-cell deployment scheme is designed to cope with the ``flash crowd traffic" resulted by a crowded event such as parades and concerts.

Due to the high altitude and the flexible deployment of the UAVs, the links between the drone-cell BSs and the ground nodes can be very reliable by mitigating the blockage effects. Therefore, the wireless fronthaul as well as the radio access of the drone-cells can employ the mmWAVE and beamforming technologies to provide high-speed connections. However, due to the limited transmit power of UAVs, the fronthaul link has a limited capacity. Therefore, the drone-cells are more suitable for the IoT services, which usually have small packets and low rate requirements. Since the coverage of drone-cells may vary according to the drone-BSs' altitude and transmission power, multi-tier heterogeneous drone-cells can be constructed. In spite of the enhanced capacity, the multi-tier architecture of drone-cells will result in severe interference problem, which requires carefully designed interference management.

\subsection{Mobile Edge Caching in AGMEN}
In traditional MEN, popular contents are usually cached proactively in the mobile edge, such as small-cell BSs for reducing the backhaul burden and service delay. However, when mobile users move outside the cell coverage areas, the cached contents may not be effectively distributed to the users. In addition, when the user handovers to a new cell, the contents she requests may not be cached, leading to an extra delay and bandwidth consumption due to the caching in the new BS or long-distance fetch from the content server. Also, in drone-cells, the limited fronthaul capacity can hardly satisfy the demands for data-craving services. To solve the problems, we proposed UAV-assisted edge caching in AGMEN.

In UAV-assisted edge caching, the contents can be directly cached in the drone-BSs, and distributed to users, or cached in the mobile devices and scheduled by the drone-BSs. For the former one, the contents can be cached during the off-peak hours, or when the UAVs are at the docking station. For the latter one, vehicles or mobile users can cache the contents they previously requested, and distribute such contents among nearby users following the scheduling of ground BSs or drone-BSs. Such UAV-assisted edge caching can bring many potential benefits. The fronthaul bandwidth of drone-cells can be saved, and therefore data-craving applications can also be served through drone-cells with low latency. In addition, the caching capability of the edge network is fully utilized while the caching strategy and content distribution decisions can be made by the drone-BSs in a flexible manner, which improves the caching and energy efficiency of AGMEN.
\begin{figure*}
\centering
\includegraphics[height=7.2cm]{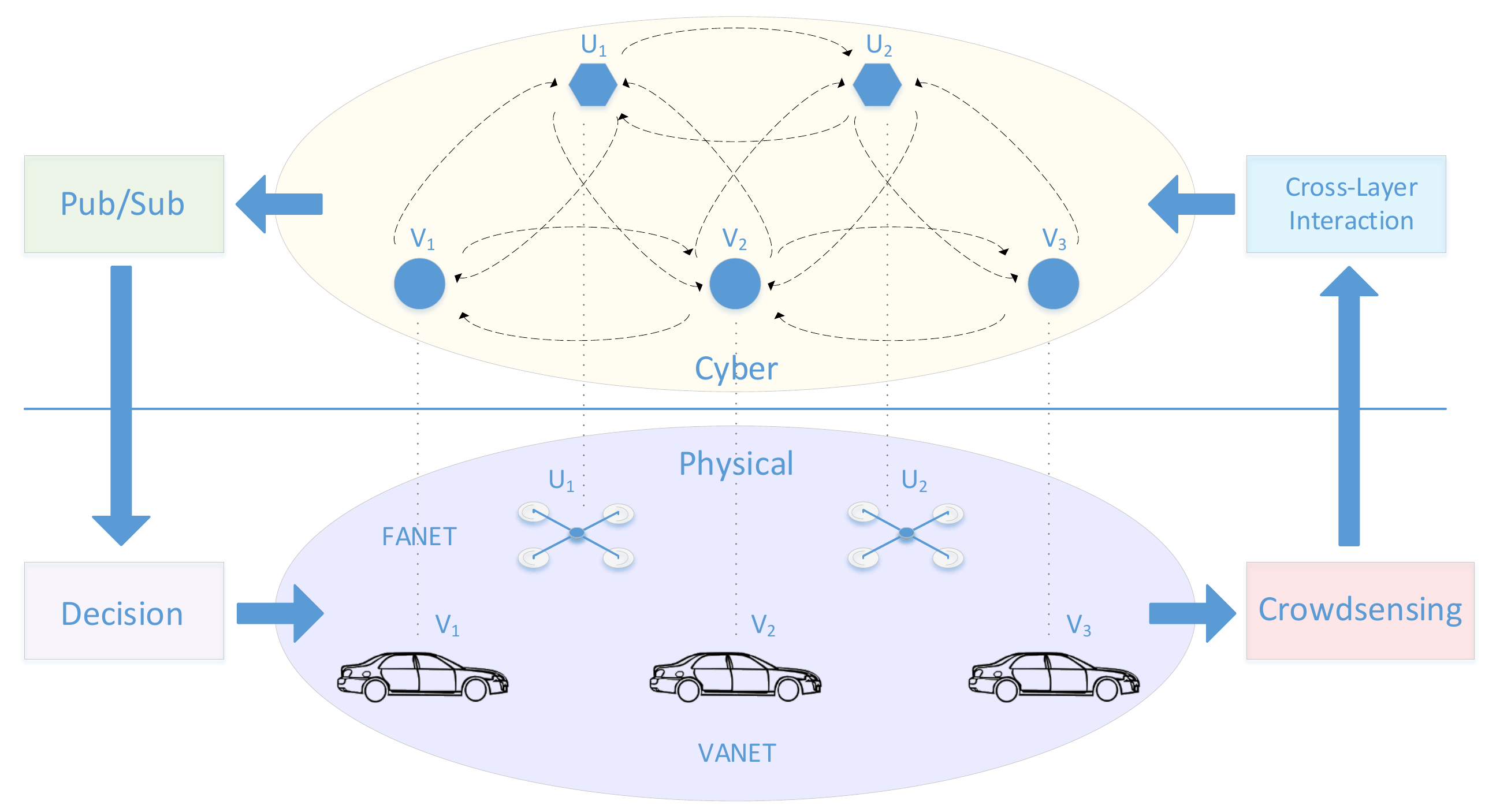}
\caption{Cyber-physical architecture of UAV-assisted crowdsensing.}
\label{fig:cp}
\end{figure*}

\subsection{Edge Computing for IoT in AGMEN}
Outfitted with IoT devices and cooperating with each other, the UAVs on the aerial network can form a flying fog computing platform, providing flexible and resilient services to IoT devices with limited processing capabilities. For example, the computationally heavy tasks, such as face recognition and VR, can be offloaded to the UAVs through A2G communications. In addition, by means of aerial-ground cooperation, the IoT devices on both layers can work together to carry out some complicated or special IoT services, such as search and rescue after disaster, large-scale crowd sensing, etc. In these applications, the UAVs can fly to the area of interest, obtain the global information from the sky, and guide the ground stations for fine-grained tasks. The ground stations, especially the self-driving cars, can utilize their abundant computing resources for processing the raw data, and transmit the interested information with reduced size to the UAVs.

One typical application of such edge computing platform is crowd surveillance where a swarm of UAVs monitor the places or events where crowds of people get together, such as sports events or parades. Instead of security guards, the UAVs can take high-quality surveillance videos and sense other valuable information using the on-board IoT devices. After collecting the data, UAVs can process it locally, or offload the computation-heavy tasks to the mobile edge servers or even mobile user devices through the drone-cell fronthaul or A2G communications. For example, face recognition in videos needs a large amount of computing resources to conduct the detection, segmentation, and recognition tasks through computer vision technologies such as deep convolutional neural network (CNN), and therefore such tasks can be offloaded to the mobile devices with advanced neural processing chips, such as Apple iPhone X, or cars equipped with AI car computers.

In another application where UAVs assist crowd sensing, UAVs can schedule and plan the sensing tasks for aerial and ground stations, and evaluate the costs through the publication/subscription (pub/sub) mechanism. A cyber-physical architecture can be applied, where in cyber dimension tasks are scheduled and sensing data is collected through communication, while in physical dimension UAVs and vehicles fulfill the sensing tasks in real world, as shown in Fig. \ref{fig:cp}. In the crowd sensing task of fine-grained road information collection for the maintenance of high-definition (HD) map for self-driving, the UAVs can fly to the areas of interest, and publish crowd sensing tasks according to the coarse evaluation of the area from high altitude. Then, the fine-grained sensing tasks can be scheduled by UAVs and sent to self-driving cars in order to efficiently collect information. Since the information collected is with significant large size (from cameras and LiDARs), the cars can employ the on-board computing resources to compare it with the current HD map, and send only the difference to the UAVs to save the bandwidth and energy.

\section{Challenges in AGMEN}
In AGMEN, the air-ground integrated network is featured by three-dimensional mobility, dynamic topology, time-varying channel condition, and frequent air-ground interactions, which together lead to difficulties in network analysis and optimization. In this section, several challenging issues in AGMEN are presented, and potential solutions are discussed.

\begin{figure*}
\centering
\includegraphics[height=7cm]{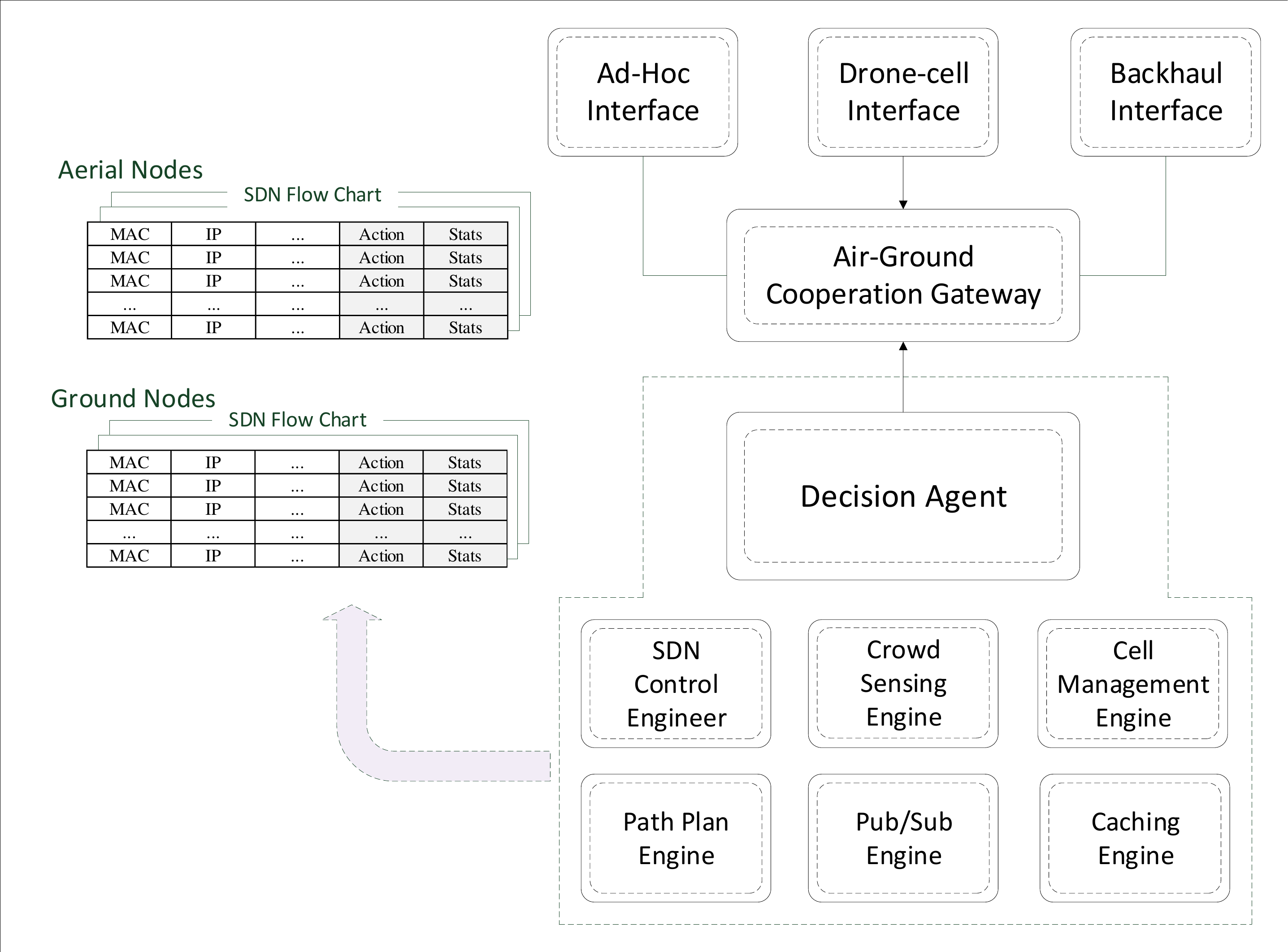}
\caption{SDN-based AGMEN management framework.}
\label{fig:sdn}
\end{figure*}

\subsection{Network Interworking}
As a typical heterogeneous network consisting of various communication nodes in both aerial and ground layer, the interworking of different network elements is a challenging problem faced by AGMEN, where two issues have to be considered:
1) the heterogeneity of devices.
Different essential network elements in AGMEN (vehicular network, HetNets, FANET) are supported by their specific communication technologies respectively. Therefore, the data exchanges among AGMEN nodes have to be conducted in a multi-protocol environment. Interfaces connecting different network elements have to be designed for seamless interactions between them; and
2) the dynamic topology.
Both VANETs and FANETs in AGMEN have dynamic topologies which degrades the quality of communication channels. Therefore, communication technologies with dedicated schemes handling the impacts of mobility are required for intra-communications within aerial and ground network, and inter-communications between the two networks in AGMEN.

\begin{figure*}
\centering
\includegraphics[height=6cm]{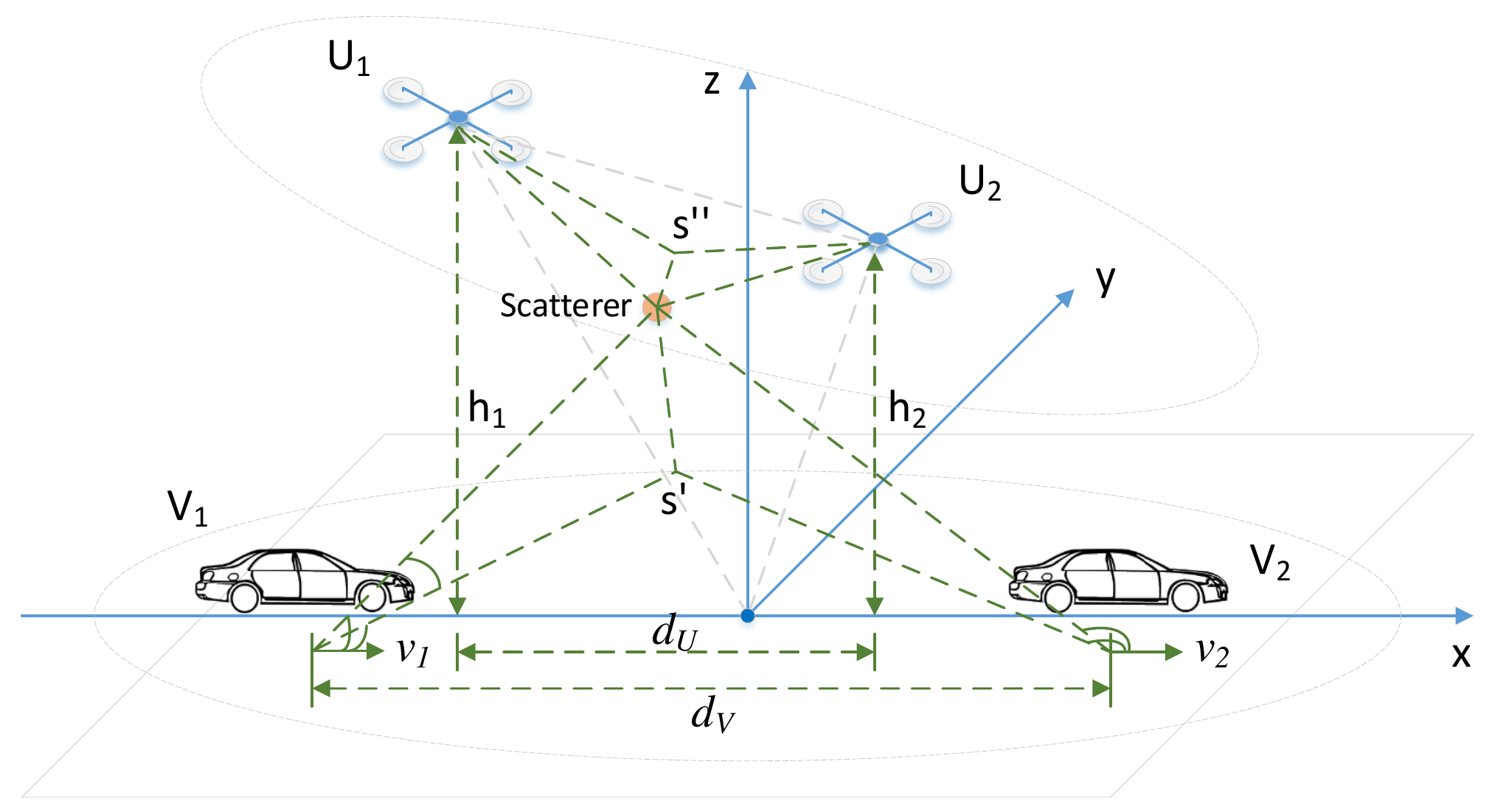}
\caption{3D GSMC modeling of AGMEN channels.}
\label{fig:gsmc}
\end{figure*}

\subsection{SDN-based Cooperative Control and Communication}
In AGMEN, dedicated control schemes are required to comprehensive control the movement of UAVs, charge/discharge behaviors, and schedule communication and computation tasks. Software defined network (SDN) refers to a network paradigms that separates the network control plane from the data plane. With a global knowledge and powerful control over the network, the SDN controller can efficiently allocate resources and functions, enhancing the flexibility, efficiency, interoperability, and reliability of the network. For example, Quan \emph{et al.} proposed an elegant crowd collaboration scheme for software defined vehicular networks, and showed the efficiency in supporting various vehicular applications through crowdsensing-based network function slicing \cite{quan2017enhancing}. To improve the efficiency on cooperative control and communication in AGMEN, SDN can be employed in the AGMEN management framework, as shown in Fig. \ref{fig:sdn}. Based on the SDN architecture, vehicles and UAVs can perform as switches and crowdsensing nodes which collect context information in a distributed way, while the BSs are controllers gathering data and making control decisions on network functions and resource allocation, which control the network behaviors through SDN flows. To reduce the traffic of control messages and improve control efficiency, selected vehicles and UAVs can perform as sub-controllers (cluster heads) to process local control requests, which corresponds to the hierarchical controller architectures in SDN.

\subsection{Cognition, Prediction and Optimization of Communication Links}
Aside from V2V and V2I communication links in traditional vehicular networks, multiple new types of wireless links are involved in AGMEN, including UAV-to-UAV (U2U), UAV-to-vehicle (U2V), and UAV-to-BS (U2B) links.
The features and QoS requirements of different links vary significantly.
For instance, the 3D mobility feature of UAVs makes antenna direction a dominating impactor for U2U communications.
Significant Doppler shifts and channel fading appear on U2V links due to the high-mobility of both UAVs and vehicles.
When UAVs act as drone-BSs, high QoS has to be guaranteed for U2B links for supporting large throughput.
Some pioneer works have been done to statistically model the A2G and U2B channels \cite{al2014optimal}. However, more detailed and dynamic modeling, predicting and optimizing of communication links in AGMEN, are still missing.

\begin{figure*}
\centering
\includegraphics[height=6cm]{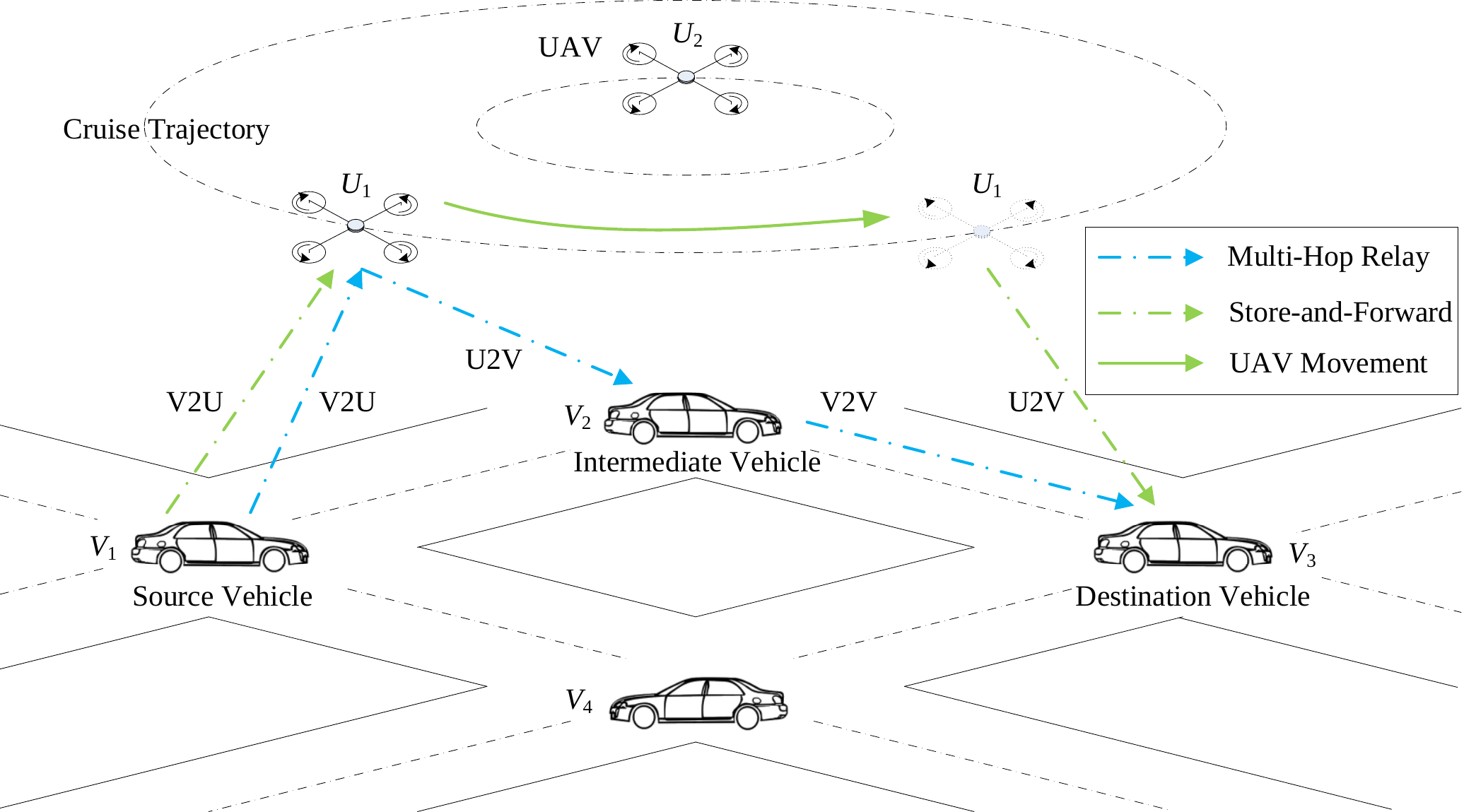}
\caption{UAV-assisted data delivery in vehicular DTN.}
\label{fig:dd}
\end{figure*}
\subsection{Evaluation Methods of AGMEN}
AGMEN is a complicated system featured by HetNets, dynamic 3D mobility, air-ground integration, and numerous applications with varied requirements. Therefore, building up a comprehensive evaluation system to simulate, test and validate the performance of AGMEN is a crucial task. Assisted by emerging technologies such as Hardware-in-the-loop (HIL) and SDN, the simulation and evaluation of AGMEN can be conducted in an integrated environment involving different simulation platforms and real systems, while the network can be easily reconstructed by the software-defined functions to reduce the system complexity. Besides, real data collected through road tests of vehicles and UAVs can be desirable sources for data analysis and simulations by means of big data mechanisms.

\section{Open Research Issues for AGMEN}
In spite of its potentials, the research on UAV-assisted AGMEN is still in its infant stage, where many key research issues are still open. In this section, we discuss the potential research issues, and highlight interesting research topics for future study.

\subsection{Modeling and Optimization of Mobile Routing}
Mobile routing is a key problem in AGMEN research, especially for the aerial network. Though the 3D mobility feature of UAVs increases the complexity of routing topology, additional distribution space is released at the same time to achieve better U2U and A2G connections. Combining with the channel state indicator (CSI) provided by channel estimation technology, SDN controllers and sub-controllers can calculate and implement the optimized routing solution. Various mathematical tools are available to model and solve mobile routing optimization problems. For instance, the mobile routing configuration and modeling problem can be described in stochastic geometry models. K-connected-center cost functions are promising tools to calculate routing constraints.

\subsection{Stochastic Optimization of Multi-Dimension AGMEN Channel}
In AGMEN, the dynamics in the aerial-ground cooperation networks leads to severe uncertainty in the wireless channels, which requires careful study to enhance the network performance. Considering the time-variant features of wireless links, geometric-based stochastic channel (GSMC) modeling technology can be employed in system recognition and coefficient estimation of A2A, A2G, and V2V channels. In GSMC modeling of AGMEN channels, the specific TX, RX, and scatters are drawn in a stochastic manner in 3D space, based on the stochastic parameters tuned for the environment scenario, as shown in Fig. \ref{fig:gsmc}. In addition, the Galerkin Projection based Pattern Downgrade method is a promising way to solve the multi-coefficient, large-scale, multi-target and high-dimensional reverse uncertainty quantify problems in AGMEN channel optimization.

\subsection{Intelligent UAV Scheduling}
In AGMEN, UAVs play a core role since they can not only provide drone-cell, caching and computing services, but also control the network functionalities such as resource allocation, task scheduling, etc. On the other hand, the energy of UAVs is constrained and high energy efficiency should be achieved. Therefore, in AGMEN, the UAV mobility and mission scheduling is a critical and challenging issue. Several important factors should be jointly considered in the UAV scheduling problem, as discussed in the following.
\begin{enumerate}
\item[$\bullet$] \textit{Joint performance optimization of drone-cell, edge caching, and edge computing}. The scheduling of the limited number of UAVs should consider the request of the mobile data traffic and IoT computing tasks. Therefore, the scheduling scheme of UAVs as drones-cells and IoT computing devices should be investigated to achieve a proper tradeoff on the performance of data traffic accommodation and edge computing efficiency.
\item[$\bullet$] \textit{Prediction of user mobility and service/content request.} The mobility and service/content requests of users can be predicted, and such valuable information can be efficiently employed in the scheduling of UAV paths in order to improve the overall network performance.
\item[$\bullet$] \textit{Energy efficiency.} Generally, the UAVs have a limited energy and have to charge when the battery drains, which could lead to the failure of network nodes. Therefore, the UAV scheduling schemes should be aware of the energy constraints, and guarantee that the network functions normally under such constraints.
\end{enumerate}

\subsection{UAV-Assisted Data Delivery}
Since UAVs can enable dynamic deployments and build line-of-sight (LoS) connections to ground nodes with high-quality links, it is a promising technology to employ UAVs to assist data distribution in AGMEN.
UAVs can mainly play two roles to enhance the data delivery in AGMEN:
1) serving as the drone-BSs, UAVs can enhance the connectivity of ground network via high-reliability and high-rate A2G links, in areas where the ground network is too sparse to set up direct communication links. Efficient transmission technologies, such as MIMO and beamforming can be utilized to further enhance the network performance. By dynamically adjusting the altitude and transmission power, the tradeoff between coverage and interference can be optimized; and
2) the mobility of UAVs can be employed to deliver data in a delay/disruption tolerant manner, where UAVs act as ``flying data mules", as shown in Fig. \ref{fig:dd}. The routing and delivery protocols can be designed based on the mobility prediction of UAVs and ground nodes.

\section{Conclusion}
\label{section: Conclusion}
In this article, we have proposed an air-ground integrated mobile edge network, where efficient UAV scheduling and air-ground cooperation are employed to jointly optimize the performance of AGMEN functions, i.e., drone-cells, edge caching and edge computing. We have described the architecture and key components of AGMEN, while the support of specific applications in IoT has been explained. The challenging issues in such an architecture have also been discussed, and potential research directions have been presented for future study.

\section*{Acknowledgment}
This work is sponsored in part by the National Natural Science Foundation of China (NSFC) under Grant No. 91638204, and the Natural Sciences and Engineering Research Council of Canada.

\bibliographystyle{IEEEtran}
\bibliography{MyRefs}

\vspace*{-1.5\baselineskip}
\begin{IEEEbiography}{Nan Cheng}
[S'12, M'16] received the Ph.D. degree from the Department of Electrical and Computer Engineering, University of Waterloo. He is currently working as a Post-doctoral fellow with the Department of Electrical and Computer Engineering, University of Toronto, and with Department of Electrical and Computer Engineering, University of Waterloo. His research interests include performance analysis and opportunistic communications for vehicular networks, unmanned aerial vehicles, and cellular traffic offloading.
\end{IEEEbiography}

\vspace*{-1.5\baselineskip}
\begin{IEEEbiography}{Wenchao Xu}
received the B.E. and M.E. degrees from Zhejiang University, Hangzhou, China, in 2008 and 2011, respectively. He is currently working toward the Ph.D. degree with the Department of Electrical and Computer Engineering, University of Waterloo, Waterloo, ON, Canada. In 2011, he joined Alcatel Lucent Shanghai Bell Co. Ltd., where he was a Software Engineer for telecom virtualization. His interests include wireless communications with emphasis on resource allocation, network modeling, and mobile data offloading.
\end{IEEEbiography}

\vspace*{-1.5\baselineskip}
\begin{IEEEbiography}{Weisen Shi}
received the B.S. degree from Tianjin University, Tianjin, China, in 2013 and received the M.S. degree from Beijing University of Posts and Telecommunications, Beijing, China, in 2016. He is currently working toward the Ph.D. degree with the Department of Electrical and Computer Engineering, University of Waterloo, Waterloo, ON, Canada. His interests include drone communication and networking, network function virtualization and vehicular networks.
\end{IEEEbiography}

\vspace*{-1.5\baselineskip}
\begin{IEEEbiography}{Yi Zhou}
received the B.S. degree in electronic engineering from the First Aeronautic Institute of Air Force, China, in 2002, and the Ph.D. degree in control system and theory from Tongji University, China, in 2011. He is currently an Associate Professor with the School of Computer and Information Engineering, Henan University, Kaifeng, China. His research interests include vehicular cyber-physical systems and multi-agent design for vehicular networks.
\end{IEEEbiography}

\vspace*{-1.5\baselineskip}
\begin{IEEEbiography}{Ning Lu}
is an assistant professor in the Dept. of Computing Science at Thompson Rivers University, Canada. He received his Ph.D. degree in Electrical and Computer Engineering from the University of Waterloo, Canada, in 2015, with the thesis supervised by Prof. Xuemin (Sherman) Shen. He received the B.E. and M.E. degrees both in electrical engineering from Tongji University, Shanghai, China. His current research interests include real-time scheduling, distributed algorithms, and reinforcement learning for wireless networks.
\end{IEEEbiography}

\vspace*{-1.5\baselineskip}
\begin{IEEEbiography}{Haibo Zhou}
[M'14, SM'18] received the Ph.D. degree in information and communication engineering from Shanghai Jiao Tong University, Shanghai, China, in 2014. From 2014 to 2017, he has worked a Post-Doctoral Fellow with the Broadband Communications Research Group, ECE Department, University of Waterloo. Currently, he is an Associate Professor with the School of Electronic Science and Engineering, Nanjing University. His research interests include resource management and protocol design in cognitive radio networks and vehicular networks.
\end{IEEEbiography}

\vspace*{-1.5\baselineskip}
\begin{IEEEbiography}{Xuemin (Sherman) Shen}
[F] is a university professor, Department of Electrical and Computer Engineering, University of Waterloo, Canada. His research focuses on resource management, wireless network security, social networks, smart grid, and vehicular ad hoc networks. He is an IEEE Fellow, an Engineering Institute of Canada Fellow, a Canadian Academy of Engineering Fellow, and a Royal Society of Canada Fellow. He was a Distinguished Lecturer of the IEEE Vehicular Technology Society and the IEEE Communications Society.
\end{IEEEbiography}

\end{document}